\documentclass[aps,preprint]{revtex4}
\usepackage{amsmath,amssymb,amsthm,mathrsfs,amsfonts,dsfont} 
\usepackage{graphicx}
\usepackage{subfigure}
\usepackage{amsbsy}
\usepackage{bm}
\usepackage{xcolor}
\usepackage{dcolumn}

\DeclareMathOperator{\Tr}{Tr}

\usepackage{algorithm}
\usepackage{algorithmicx}
\usepackage{algpseudocode}

\usepackage{amsmath}
\usepackage{array}
\usepackage{booktabs}

\def\id{\mathds{1}}

\def\e{\mathrm{e}}
\def\i{\mathrm{i}}

\begin{document}

\title{Freezing chaos without synaptic plasticity}
\author{Weizhong Huang$^{1}$}
\author{Haiping Huang$^{1,2}$}
\email{huanghp7@mail.sysu.edu.cn}
\affiliation{$^{1}$PMI Lab, School of Physics,
Sun Yat-sen University, Guangzhou 510275, People's Republic of China}
\affiliation{$^{2}$Guangdong Provincial Key Laboratory of Magnetoelectric Physics and Devices,
Sun Yat-sen University, Guangzhou 510275, People's Republic of China}
\date{\today}

\begin{abstract}
	Chaos is ubiquitous in high-dimensional neural dynamics. A strong chaotic fluctuation may be harmful to information processing. A traditional way to mitigate this issue is to introduce Hebbian
	plasticity, which can stabilize the dynamics. Here, we introduce another distinct way without synaptic plasticity. An Onsager reaction term due to the feedback of the neuron itself is added to the vanilla recurrent dynamics, making the driving force a gradient form. The original unstable fixed points supporting the chaotic fluctuation can then be approached by further decreasing the kinetic energy of the dynamics. We show that this freezing effect also holds in more biologically realistic networks, such as those composed of excitatory and inhibitory neurons. The gradient dynamics are also useful for computational tasks such as recalling or predicting external time-dependent stimuli.
\end{abstract}

 \maketitle

\section{Introduction}
Recurrent neural networks (RNNs) with asymmetric couplings between neurons are ideal candidates for theoretical studies of non-equilibrium dynamics~\cite{Chaos-1988,Pham-2025,Qiu-2024}, linking machine learning, statistical physics and neural computation~\cite{Huang-2024,Kadmon-2024}. We consider continuous neural dynamics and the couplings between pairs of neurons are independently drawn from a normal distribution with zero mean and a scaled variance. Previous studies confirmed that increasing the coupling variance will trigger a chaos transition in a continuous way~\cite{Chaos-1988,Qiu-2024}. A recent study pointed out that the chaotic fluctuation can be frozen by a Hebbian synaptic plasticity added to the original random couplings~\cite{Clark-2024}. However, this may not be a unique way to manipulate the chaos. Our recent work provided an optimization perspective on the high-dimensional chaos in the RNN~\cite{Qiu-2024},  which further demonstrated that the non-gradient force for the dynamics can be decomposed into a gradient one and an Onsager reaction (OR) type whose value depends on the dynamics speed in turn.  The OR force can be controlled, e.g., switched on or off, which yields a large impact on the ongoing dynamics of the original system in the chaotic regime. It remains nevertheless unclear what is the emergent property of this force decomposition. This is the central question to be addressed in this work.

In this paper, we fully explore the collective properties of the random recurrent neural networks with controlled forces. Our main contributions are three-fold. First, we discover an additional way of freezing chaos, especially without synaptic plasticity, and in other words, by turning on the OR term. It is expected that as an additional white noise is added to the potential system (because of gradient dynamics) where the OR term is turned on such that a gradient force is guaranteed, the original dynamics can be approximated by making the noise very small in magnitude. This leads to our second contribution that the gradient system exhibits three phases: null-activity, nontrivial fixed points yet with vanishing maximal Lyapunov exponents, and the chaotic phase (at a larger coupling variance than that in the standard non-gradient network dynamics). These properties carry over to the biologically plausible networks of excitatory and inhibitory neurons. Third, 
the memory or prediction curve is also studied, showing that the potential system at a large coupling variance matches the original non-gradient one in behavior. In particular, the intermediate slow activity phase could be useful in computation, e.g., maintaining a working memory of a generated mental picture.

\section{Model setting}\label{prob}
\subsection{Vanilla RNN dynamics }
We consider a recurrent neural network composed of $N$ fully-connected neurons. 
The state of each neuron in time $t$ is characterized by the synaptic current $x_i(t)$ ($i= 1,\ldots,N$), which obeys the following first-order differential
 dynamics equation:
\begin{equation}\label{rnn}
 \frac{{dx_i}}{{dt}} =  - x_i + \sum\limits_{j = 1}^N {J_{ij}} \phi _j(t),
\end{equation}
where $\phi _j(t) = \tanh [x_j(t)]$ transforms synaptic current to firing rate. We denote $x_{i} (t)$ as a time-dependent activity, which is shown in the result section unless otherwise stated. Each element $J_{ij}$ (from neuron $j$ to $i$) of the connection matrix is drawn from an independent Gaussian distribution as
\begin{equation}
   J_{ij} \sim \begin{cases}\mathcal{N}\left(0, \frac{g^2}{N}\right) & \text { for } i \neq j \\ 0 & \text { for } i=j\end{cases} .
\end{equation} 
Increasing the value of $g$ (the gain parameter) changes the dynamics phase from a global stable fixed point to a proliferation of an exponential number of unstable fixed points (deterministic chaos)~\cite{PRL-2013}. This fact identifies a critical continuous chaos transition point $g_c=1$~\cite{Qiu-2024}. Equation~\eqref{rnn} is a rate dynamics abstraction of a more biological setting such as leaky integrated firing model~\cite{Huang-2022}. We later consider the biologically realistic scenario of excitatory-inhibitory neural networks. In the following analysis, we also denote the non-gradient force as $\mathbf{f}=-\mathbf{x}+\mathbf{J}\phi(\mathbf{x})$.

\subsection{Potential dynamics description}
Recent studies showed that if the kinetic energy (unit mass) is defined, one can turn the original dynamics [Eq.~\eqref{rnn}] into an optimization problem of searching for the slow dynamics with minimal kinetic energy defined below:
\begin{equation}\label{ke}
E_{\rm k}=\frac{1}{2}\sum_{i}\dot{x}_{i}^{2},
\end{equation}
where $\dot{x}_{i}$ represents the velocity in Eq.~\eqref{rnn}. One motivation is the observation that the kinetic energy is decreasing over running Eq.~\eqref{rnn}. Therefore, we can design an alternate Langevin dynamics:
\begin{equation}\label{sde}
    \frac{\mathrm{d}\mathbf{x}}{\mathrm{d}t}=-\nabla_\mathbf{x}E_{\rm k}(\mathbf{x})+\sqrt{2T}\boldsymbol{\epsilon},
\end{equation}
where  $\boldsymbol{\epsilon}$ is a time-dependent white noise whose statistics is given by $\langle\epsilon_{i}(t)\rangle=0,\langle\epsilon_{i}(t)\epsilon_{j}(s)\rangle=\delta_{ij}\delta(t-s)$. The temperature $T$ adjusts the noise intensity. 

We next write the gradient over the kinetic energy explicitly as follows,
\begin{equation}\label{grad}
-\frac{\partial E_{\rm k}}{\partial x_{i}}=-x_{i}+h_{i}-\phi^{\prime}(x_{i})\sum_{j\neq i}J_{ji}(h_{j}-x_{j}),
\end{equation}
where the local field $h_{i}\equiv\sum_{j:j\neq i}J_{ij}\phi(x_{j})$. It is clear that the third term in Eq.~\eqref{grad} is an Onsager reaction term in dynamics because this term represents how the neuron $i$ impacts its neighbors through the outgoing connections $J_{ji}$ and then affects the $i$-component of the force as feedback together with the velocity of neighboring neurons (see a similar study but in the context of kinetic Ising models~\cite{Huang-2014}). We thus call this term an Onsager reaction term, which plays a vital role in slowing down the chaotic fluctuation and making neural activity a suitable candidate for working memory. 

\subsection{Neural force decomposition}
We also notice that the first two terms in Eq.~\eqref{grad} are the original force, and thus we obtain the following identity:
\begin{equation}\label{forceeq}
\mathbf{f}=-\nabla_{\mathbf{x}}E_{\rm k}+\phi'(\mathbf{x})\odot\mathbf{J}^{\top}\mathbf{f},
\end{equation}
where $\odot$ is an elementwise multiplication. From Eq.~\eqref{forceeq}, we can re-express the force as 
\begin{equation}
\mathbf{f}=-\mathbf{\tilde{J}}^{-1}\nabla_{\mathbf{x}}E_{\rm k},
\end{equation}
where $\tilde{J}_{ij}\equiv J_{ji}\left[\delta_{ji}-\phi'(x_{i})(1-\delta_{ji})\right]$, and \textit{only in this formula} we assume $J_{ii}=1$. Therefore, we managed to write the non-gradient force as a linear combination of gradient force, where the linear coefficient depends not only on the original asymmetric random coupling but also on neural activity. Our construction thus gets rid of computational challenges to determine the exact form of the decomposition proposed in previous works~\cite{AP-2004,WJ-2013}, where a stochastic differential equation is studied and for high-dimensional systems studied in our work, it becomes numerically challenging to solve equations for the linear combination. 

\subsection{Regulated neural dynamics}
Inspired by the above analysis, we add a multiplicative factor $\gamma(t)$ to the OR term, and the dynamics read:
\begin{equation}\label{rnn2}
\dot{x}_{i}=-x_{i}+h_{i}-\gamma\phi^{\prime}(x_{i})\sum_{j\neq i}J_{ji}(h_{j}-x_{j})+\sqrt{2T}\epsilon_i.
\end{equation}
The dynamics control parameter $\gamma(t)$ can only take two values---$0$ and $1$. When $\gamma(t)=1$, the dynamics enter a working memory session, while the dynamics are chaotic provided that $\gamma(t)=0$.

To carry out simulations, we first adopt the simple Euler scheme to discretize the continuous dynamics
equation in Eq.~\eqref{rnn2} as follows:
\begin{equation}
x_{i}^{t+\Delta t}-x_{i}^{t}=\Delta t\left[-x_{i}^{t}+h_{i}^{t}-\gamma^{t}\phi^{\prime}(x_{i}^{t})\sum_{j\neq i}J_{ji}(h_{j}^{t}-x_{j}^{t}) \right]+\sqrt{2T\Delta t}\epsilon_{i}(t),
\end{equation}
where $\Delta t$ is a small time increment ($\Delta t=0.01$ in our following experiments), $h_{i}^{t}=\sum_{j\neq i}J_{ij}\phi(x_{j}^{t})$, and $\epsilon_{i}(t)$ is drawn independently from the standard
normal distribution for each $i$ and $t$~\cite{Yu-2025}.
At any intermediate state $\mathbf{x}$ (such as a fixed-point), one can evaluate the Jacobian or stability matrix $\mathbf{D}$ of the modified dynamics in Eq.~\eqref{rnn2} as follows:
\begin{subequations}
\begin{align}
    D_{ij}&=\frac{\partial f_i}{\partial x_j}=J_{ij}\phi^{\prime}(x_j)-\gamma\phi^{\prime}(x_i)\sum_kJ_{ki}J_{kj}\phi^{\prime}(x_j)+\gamma\phi^{\prime}(x_i)J_{ji}, \quad\forall i\neq j\\
    D_{ii}&=\frac{\partial f_i}{\partial x_i}=-1-\gamma\phi^{\prime\prime}(x_i)\sum_kJ_{ki}(h_k-x_k)-\gamma\phi^{\prime}(x_i)\sum_kJ_{ki}^2\phi^{\prime}(x_i),
\end{align}
\end{subequations}
where we omit the time dependence for relevant state variables for compactness. The eigenvalue distribution of the stability matrix determines the linear stability of the neural state where the stability analysis is carried out. Note that $\nabla\cdot\mathbf{f}=\Tr(\mathbf{D})$ measures how dissipative the neural dynamics are. The force $\mathbf{f}$ takes different forms for vanilla and regulated dynamics.

\begin{figure}
\centering
\includegraphics[width=\textwidth]{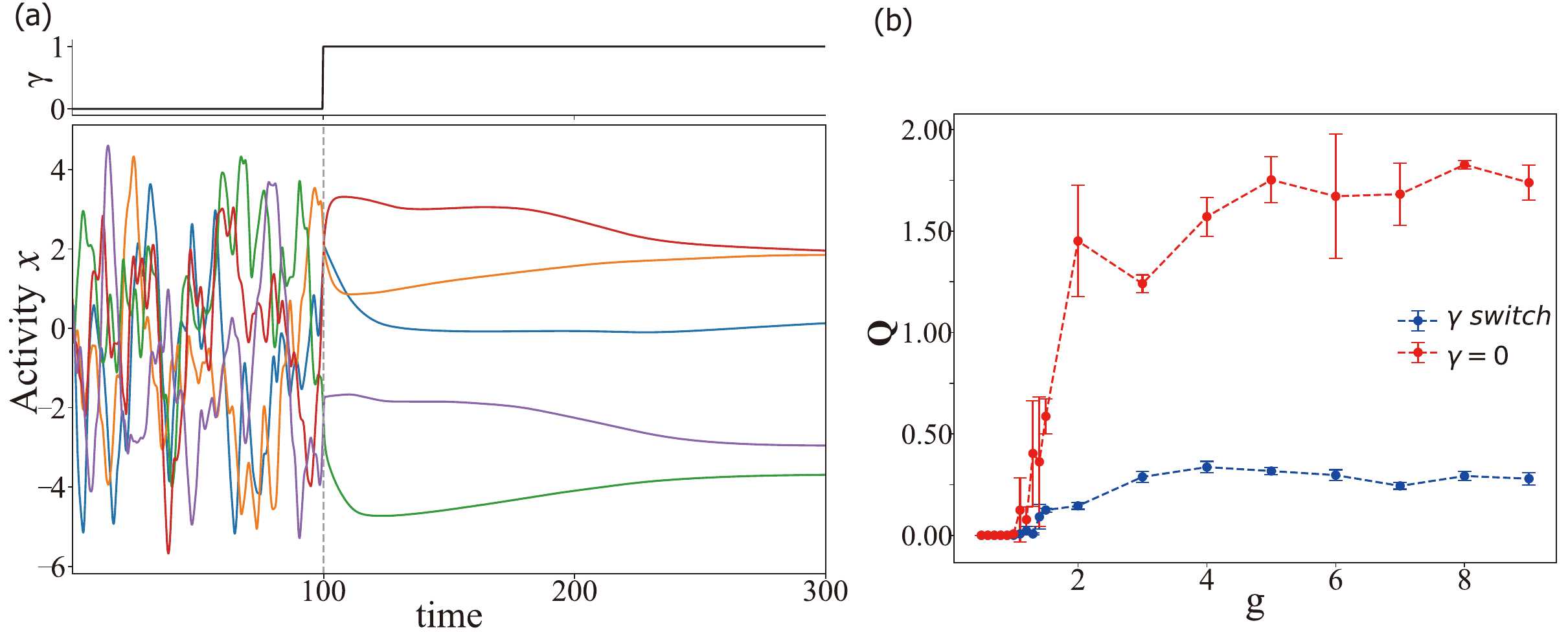}
\caption{Dynamics-freezing effects of the OR-regulated dynamics. (a) Dynamics with $T=0$, $g=3$, and $N=1000$. $\gamma$ jumps from $0$ to $1$ at the
 time point $100$. (b) The slowness-measurement $Q$ [between $t=99.99$ and $t'=300$, see (a)] changes with $g$ for both vanilla RNN and OR-regulated dynamics. Five individual trials of experiments are considered for the error bars in the plot.}
\label{fig1}
\end{figure}

\section{Results}
In this section, we thoroughly explore the collective properties of the OR term in the Langevian dynamics, in comparison with the original non-gradient dynamics.
The following results are based on the OR-regulated dynamics [see Eq.~\eqref{rnn2}]. 

\subsection{Freezing chaos without synaptic plasticity}
We find that running the OR-regulated dynamics yields a slowing down of the dynamics after $\gamma(t)$ is switched from zero to one. This is in essence that the chaotic 
fluctuations are strongly suppressed by turning on the OR term, in contrast to the way of introducing a Hebbian term to the random synaptic strength~\cite{Clark-2024}. In our dynamics 
[see Eq.~\eqref{rnn2}], the synaptic structure is still maintained without any plasticity. However, after the $\gamma$-switch, the dynamics become gradient with the kinetic energy as the underlying potential. Therefore, we propose the following parameter to characterize the slowness induced by the switch.
\begin{equation}
    Q(t,t')=\langle\|\phi(\mathbf{x}(t))-\phi(\mathbf{x}_{t^{\prime}})\|_2^2\rangle,
\end{equation}
where $t$ denotes the time point when $\gamma$ is turned on (or just one time-step before), while $t'$ is the time point $t+L$ ($L$ is the temporal separation), and the average is taken over many different
realizations of the network coupling (ensemble average).  

Figure~\ref{fig1} shows that after the onset of chaos at $g=1$, the chaotic dynamics can be frozen with the OR term. The chaotic fluctuations can be significantly suppressed. As a result, the dynamics become much more slowly, showing the great computational benefit as a working memory. The state distribution during the working memory is well-defined and given by
the canonical ensemble in statistical mechanics $P(\mathbf{x})\propto e^{-\beta E_{\rm k}(\mathbf{x})}$~\cite{Qiu-2024}, where $\beta$ is an inverse temperature controlling how noisy the dynamics are, e.g., in the Langevin dynamics [see Eq.~\eqref{sde}].

\subsection{Three phases in the potential system}
\begin{figure}
\centering
\includegraphics[width=0.9\textwidth]{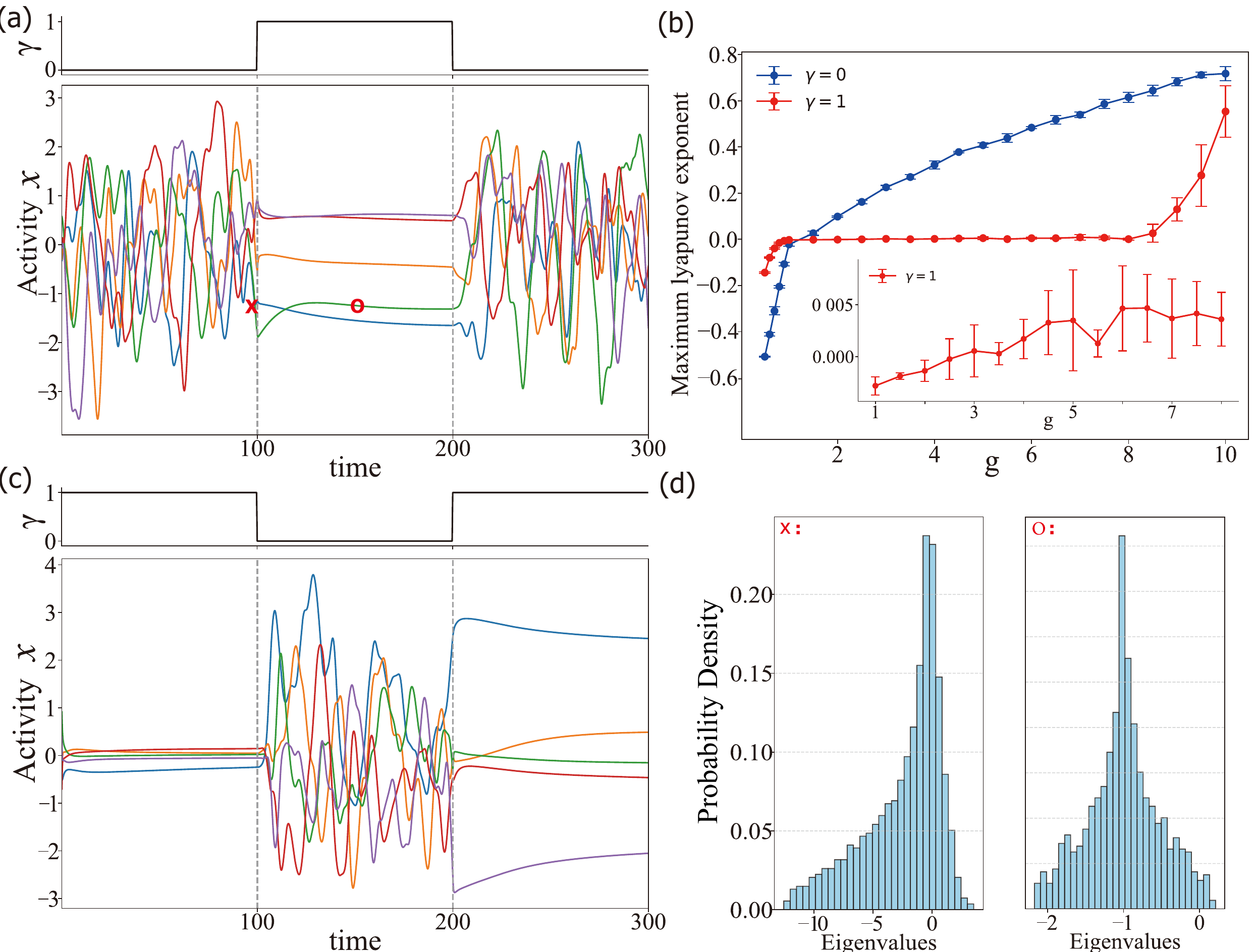}
\caption{The OR-regulated dynamics and the maximal Lyapunov exponents. (a) Dynamics with $T=0$, $g=2$ and $N=1000$. $\gamma$ goes from zero to one and back to zero over time. (b) Maximal Lyapunov exponents versus synaptic gain parameter $g$ for both vanilla and regulated dynamics ($T=0$). The exponents are averaged over five independent estimates (see details in Appendix~\ref{app-a} with an initial deviation $\delta=10^{-5}$). The inset shows an enlarged view of the intermediate phase (the exponents are close to zero). (c) Dynamics with the same model parameters as (a).
 $\gamma$ goes from one to zero and back to one over time. (d) The eigenvalue spectrum of the Jacobian matrix at the point marked by x and the point marked by o in (a).}
\label{fig2}
\end{figure}

\begin{figure}
\centering
\includegraphics[width=0.8\textwidth]{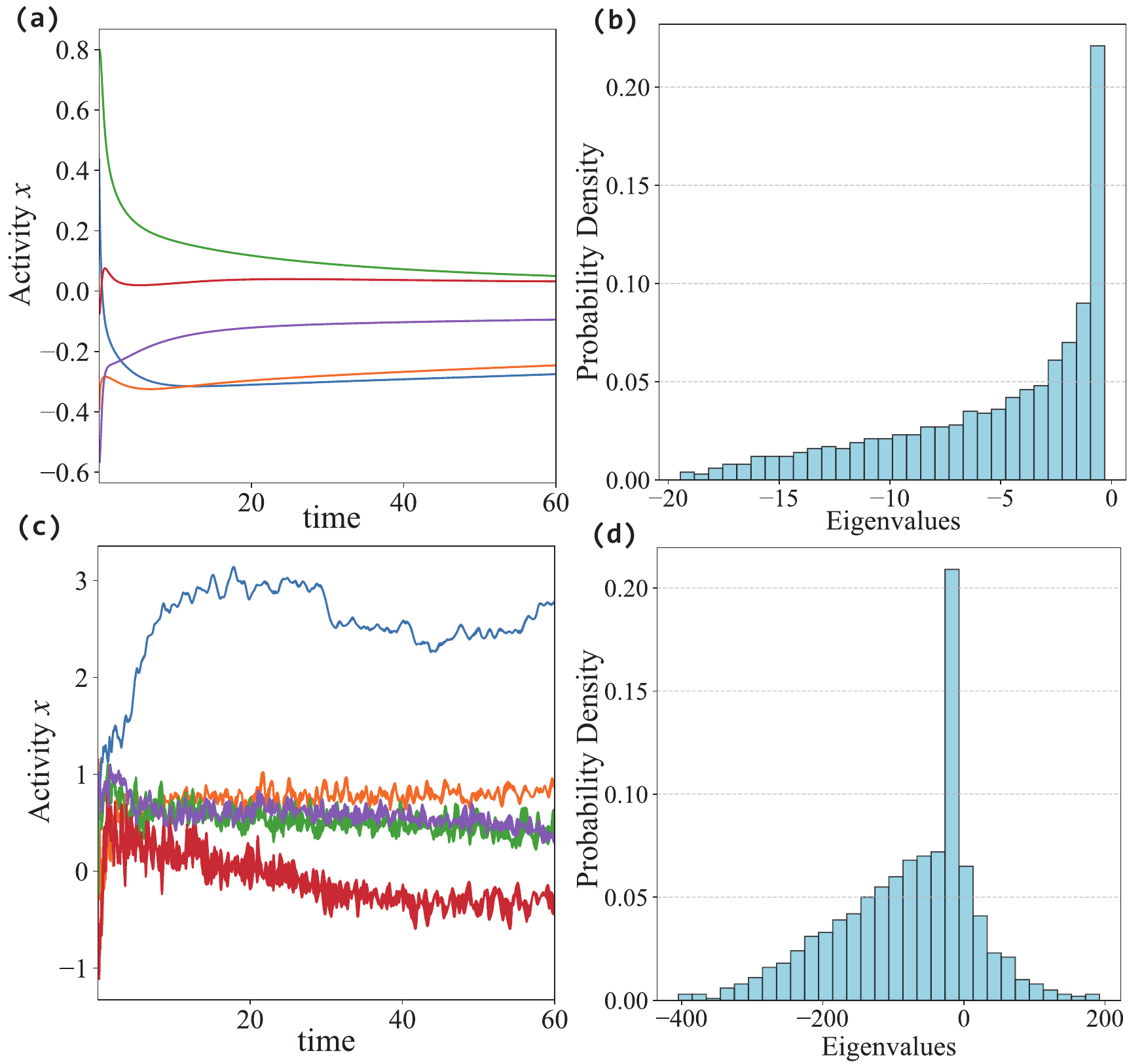}
\caption{The dynamics with $\gamma=1$ and the eigenvalue distribution of the Jacobian matrix. (a) Dynamics with $T=0$, $g=2$ and $N=1000$. (b) The eigenvalue spectrum of the Jacobian matrix at the last time point in (a). (c) Dynamics with the same model parameters as (a) but $g=10$.
  (d) The eigenvalue spectrum of the Jacobian matrix at the last time point in (c).}
\label{fig3}
\end{figure}

It is interesting to show that there appear three regimes for the potential dynamics ($\gamma=1$) in terms of the maximal Lyapunov exponents [Fig.~\ref{fig2} (b)]. We use the orbit-separation
method to evaluate the maximal Lyapunov exponents~\cite{JC-2003}. Technical details are given in Appendix~\ref{app-a}. In the vanilla dynamics [see Eq.~\eqref{rnn}], the maximal Lyapunov exponent gets positive at $g=1$, implying the onset of chaos. However, in the potential dynamics, at $g=1$, the exponent gets a value close to zero but not significantly positive, while this regime covers a very wide range of $g$. Only after $g\simeq 8$, the exponent grows rapidly, displaying a chaotic fluctuation in the potential dynamics. The dynamics flow to chaotic attractors. In the intermediate regime, the dynamics change very slowly, having the computational benefit of freezing the chaos in the vanilla dynamics. This freezing can be simply realized by turning on the OR term. As we shall show below, this regime is not suitable for retrospective and prospective computations such as recalling previous stimuli or forecasting a future event. 
However, in the intermediate regime, the dynamics can be maintained with a slow speed and can also be released to a chaotic fluctuation [see Fig.~\ref{fig2} (a, c)]. In particular, the eigen-density of the Jacobian matrix shows that the potential dynamics are much more stable [see Fig.~\ref{fig2} (d)], coinciding with our expectation of freezing the chaos with the OR term.

\begin{figure}
\centering
\includegraphics[width=0.8\textwidth]{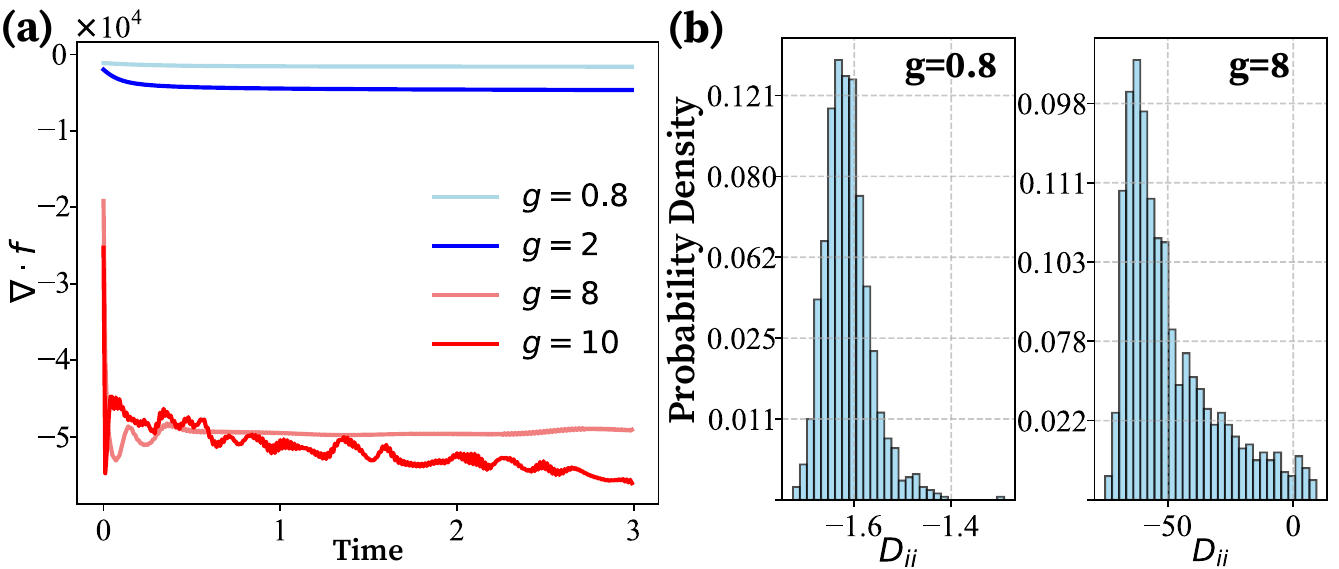}
\caption{Divergence of the driving force in the OR-regulated dynamics. $N=1\,000$, and four different values of $g$ are considered. (a) Evolution of divergence. (b) Distribution of the diagonal of the stability matrix.}
\label{fig4a}
\end{figure}

Next, we explore how different it is between the marginal stable and chaotic regimes in terms of the Jacobian matrix. Figure~\ref{fig3} shows that when $g>8$, the 
potential dynamics display a fraction of positive eigenvalues in the corresponding Jacobian matrix. Figure~\ref{fig4a} further shows how dissipative the potential dynamics are.
The negative divergence of the driving forces implies a non-uniform contraction of the phase-space volume~\cite{chaos-2009}, especially for a large value of $g$ and at an earlier stage of the dynamics.
\begin{figure}
\centering
\includegraphics[width=0.8\textwidth]{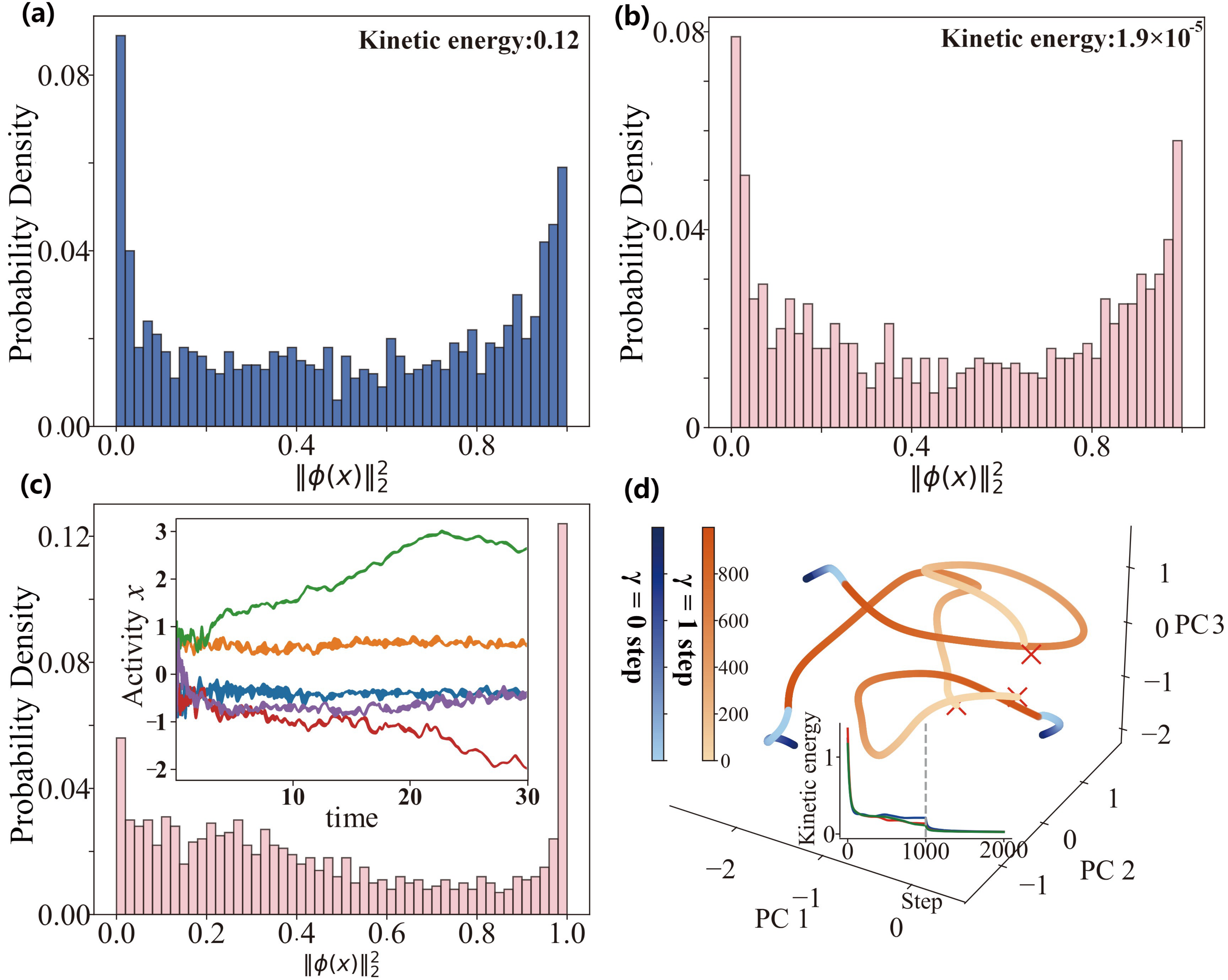}
\caption{The dynamics statistics for the OR-regulated dynamics and the associated low-dimensional projection. (a) The $\ell_2$ norm of the activity before switching with $g=2$ and $N=1000$. (b) The $\ell_2$ norm of the activity after switching with $g=2$ and $N=1000$. (c) The $\ell_2$ norm of the activity with $\gamma=1$ (gradient dynamics), $g=10$ and $N=1000$.
  (d) The three-dimensional projection of the neural dynamics where three typical trajectories are shown and the OR term is only on in the later stage of the dynamics. The first three principal components explain about $85.67\%$ of the total variance in the trajectory data. In the data, the number of trajectories $P=100$, and the length is specified by $L=2\,000$, and $N=1\,000$. The cross symbol indicates the starting point of the trajectory. The color bars indicate the flow of time for the dynamics (darker colors mean later stages). The onset of switching (from $\gamma=0$ to $\gamma=1$) is indicated by a change of the color type. The inset shows
  how the kinetic energy decreases over time for the three trajectories shown in the main plot, and the dashed line indicates the time when the OR term is on.}
\label{fig4}
\end{figure}
\subsection{Low-dimensional projection of the neural activity}
We first show the statistics of the stationary activity in the chaotic and the OR-regulated regimes. Figure~\ref{fig4} (a,b) shows that the kinetic energy 
in the chaotic regime before switching is higher than that after switching. However, the distribution of the activity norm displays a similar pattern of a U shape.
This shows that the dynamics after switching (despite its gradient nature) inherit the activity statistics of the chaotic regime (supported by the small distance $Q$ as well in Fig.~\ref{fig1}).  As $g$ grows, the distribution profile of the activity becomes a U shape as well for the gradient dynamics with the OR-regulated term [Fig.~\ref{fig4} (c)]. 

To have an intuitive picture, we try to carry out a principal component analysis (PCA) on the collected dynamics trajectories $\{\mathbf{\underline{x}}^\mu\}_{\mu=1}^{P}$, which are stacked into a matrix $\mathbf{X}\in\mathbb{R}^{PL\times N}$.
Note that $\mathbf{\underline{x}}=\{\mathbf{x}(1),\mathbf{x}(2),\ldots,\mathbf{x}(L)\}$, $P$ is the number of individual trajectories, and $L$ is the length of the trajectory. Then, we can compute the equal-time covariance:
\begin{equation}
    C_{ij}=\langle[x_i(\hat{t})-\langle x_i(\hat{t})\rangle][x_j(\hat{t})-\langle x_j(\hat{t})\rangle]\rangle,
\end{equation}
where $\hat{t}=1,\ldots,PL$, and $\left\langle\cdot\right\rangle$ denotes the temporal average.
We then perform a spectral decomposition of the covariance matrix $\mathbf{C}$ as $\boldsymbol{\Phi}\boldsymbol{\Sigma}{\boldsymbol \Phi}^{-1}$. Thus the dynamics of $\mathbf{x}(t)$ can be decomposed into the following form:
\begin{equation}
    \mathbf{x}(\hat{t})=\langle\mathbf{x}(\hat{t})\rangle+\sum_{i=1}^{N}v_{i}(\hat{t})\boldsymbol{\phi}_{i},
\end{equation}
where $v_{i}$ denotes the projection of $\mathbf{x}(\hat{t})$ along the $i$th PC direction $\boldsymbol{\phi}_{i}\mathrm{~with~}\boldsymbol{\phi}_{i}\cdot\boldsymbol{\phi}_{j}=\delta_{ij}$. In other words, $\mathbf{\underline{v}}=(\mathbf{\underline{x}}-\langle\mathbf{\underline{x}}\rangle)\boldsymbol{\Phi}$. If we collect only the first three orthogonal bases ${\boldsymbol{\phi_i}}\ (i=1,2,3)$ to form a subspace with the basis $\tilde{\boldsymbol{\Phi}}$, we would then have the three-dimensional projection of the dynamics as $\mathbf{\tilde{\underline{v}}}=(\mathbf{\underline{x}}-\langle\mathbf{\underline{x}}\rangle)\tilde{\boldsymbol{\Phi}}$.
A few typical neural dynamics trajectories are shown in Fig.~\ref{fig4} (d). With the switching (turning on the OR term), the original unstable fixed points in the chaotic regime can be achieved with a very small kinetic energy.

\subsection{Networks of excitatory and inhibitory neurons}
\begin{figure}
\centering
\includegraphics[width=\textwidth]{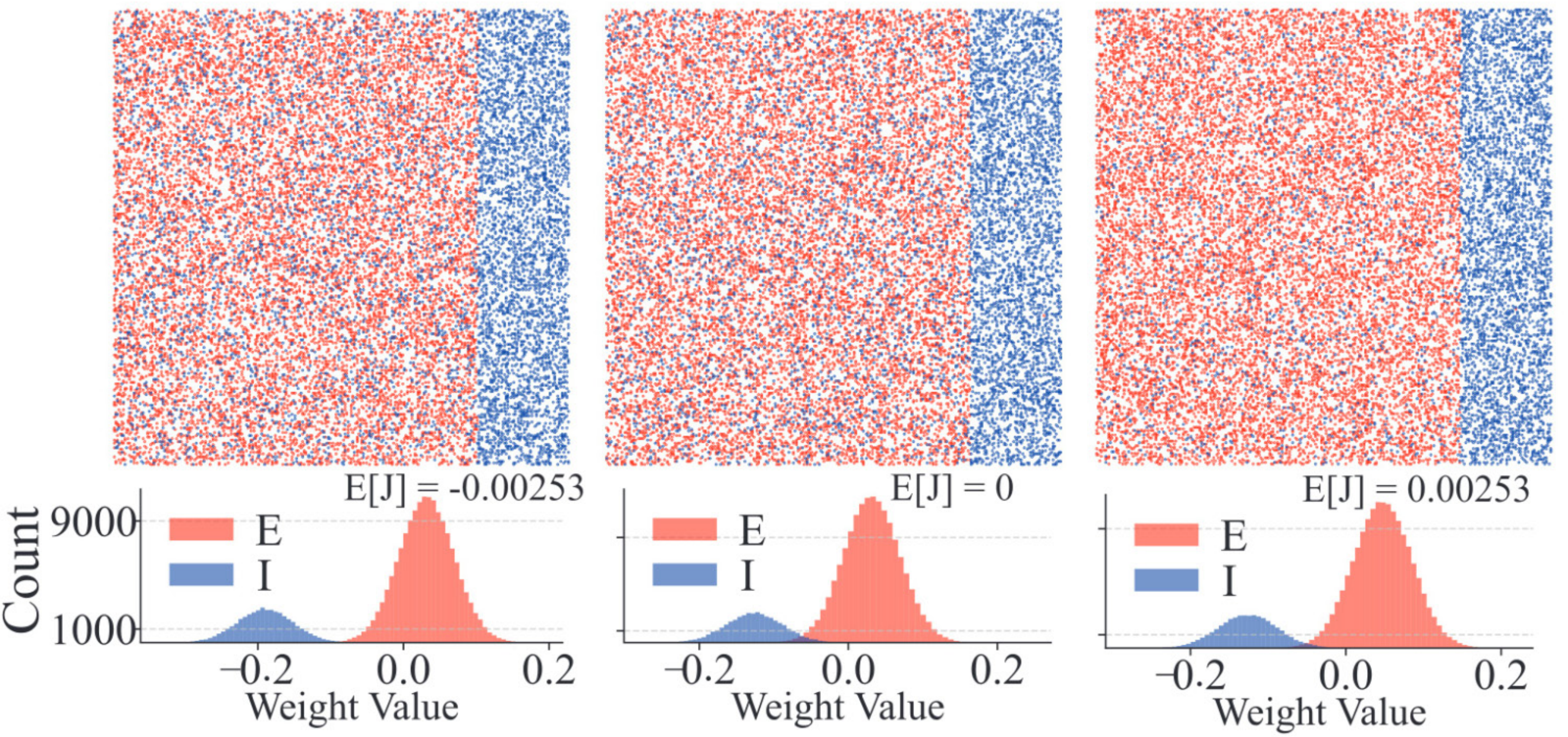}
\caption{Topology of EI neural networks ($N=1\,000$). (Left)  Inhibitory dominated network with $\mathbb{E}[J]<0$. Parameters: $\sigma_\e=1.2$, $\sigma_\i=1.2$, $\mu_\e=1$, $\mu_\i=-6$, $f=0.8$, and $\alpha=0.2$. (Middle) Structurally balanced network with vanishing
mean. Parameters: $\sigma_\e=1.2$, $\sigma_\i=1.2$, $\mu_\e=1$, $\mu_\i=-4$, $f=0.8$, and $\alpha=0.2$. (Right) Excitatory dominated network with positive mean. Parameters: $\sigma_\e=1.2$, $\sigma_\i=1.2$, $\mu_\e=1.5$, $\mu_\i=-4$, $f=0.8$, and $\alpha=0.2$. } 
\label{fig5}
\end{figure}

In this section, we do the above analysis in a more biologically realistic setting, i.e., we distinguish the neurons into two types---excitatory or
inhibitory neurons. For an excitatory neuron, the outgoing synaptic strength takes a positive value, while the outgoing synaptic strength takes 
a negative value for an inhibitory neuron. This is also called the well-known Dale's law~\cite{ND-2014}. In this setting, the coupling matrix can be expressed in the form
of blocks:
\begin{equation}
\mathbf{J}=\begin{pmatrix}\mathbf{J}^{\mathrm{EE}}&\mathbf{J}^{\mathrm{EI}}\\\mathbf{J}^{\mathrm{IE}}&\mathbf{J}^{\mathrm{II}}\end{pmatrix},
\end{equation}
where $\mathbf{J}^{\rm EE}$ indicates the connection structure from excitatory neurons to excitatory neurons, and others bear a similar meaning.
In addition, we assume this matrix is actually a sparse matrix, as in a biological circuit, a neuron can not be connected to all neurons. We apply the following
way to construct such sparse connection structure~\cite{PRR-2023}:
\begin{equation}
    \mathbf{J}=\mathbf{S}\odot(\mathbf{A}\mathbf{D}+\mathbf{M}),
\end{equation}
in which $\mathbf{S}$ denotes a sparse matrix whose entry takes a value of one with the probability $\alpha$ and zero with the probability $(1-\alpha)$,
 $\odot$ denotes the element-wise product, $\mathbf{A}$ is a random matrix in which each element is an i.i.d. Gaussian real value, $\mathbf{D}$ is
  a diagonal matrix of excitatory and inhibitory variances (weight dispersion) specified as follows,
\begin{equation}
   \mathbf{ D}=\mathrm{diag}(\underbrace{\tilde{\sigma}_\mathrm{e},\ldots,\tilde{\sigma}_\mathrm{e}}_{Nf\mathrm{~times}},\underbrace{\tilde{\sigma}_\mathrm{i},\ldots,\tilde{\sigma}_\mathrm{i}}_{N(1-f)\mathrm{~times}}),
\end{equation}
where $f$ is the fraction of excitatory neurons in the circuit.
$\mathbf{M}=\mathbf{u}\mathbf{v}^\top$ is an outer product matrix of population means, where
\begin{equation}
    \mathbf{u}=(1,\ldots,1)^{\top},\quad \mathbf{v}=(\underbrace{\tilde{\mu}_\mathrm{e},\ldots,\tilde{\mu}_\mathrm{e}}_{Nf\mathrm{~times}},\underbrace{\tilde{\mu}_\mathrm{i},\ldots,\tilde{\mu}_\mathrm{i}}_{N(1-f)\mathrm{~times}}).
\end{equation}
To avoid the dependence on the system size $N$ in particular in the limit of large system size, we rescale the mean and standard deviation as
 $\tilde{\mu}=\frac{\mu}{\sqrt{N}}$, and $\tilde{\sigma}=\frac{\sigma}{\sqrt{N}}$ in the following analysis.  We also assume $\sigma_{\rm e}=\sigma_{\rm i}=\sigma$, and this single real positive value $\sigma$ can be thought of as a control
parameter relaxing the Dale's law (see Fig.~\ref{fig5} for an example).

\begin{figure}
\centering
\includegraphics[width=\textwidth]{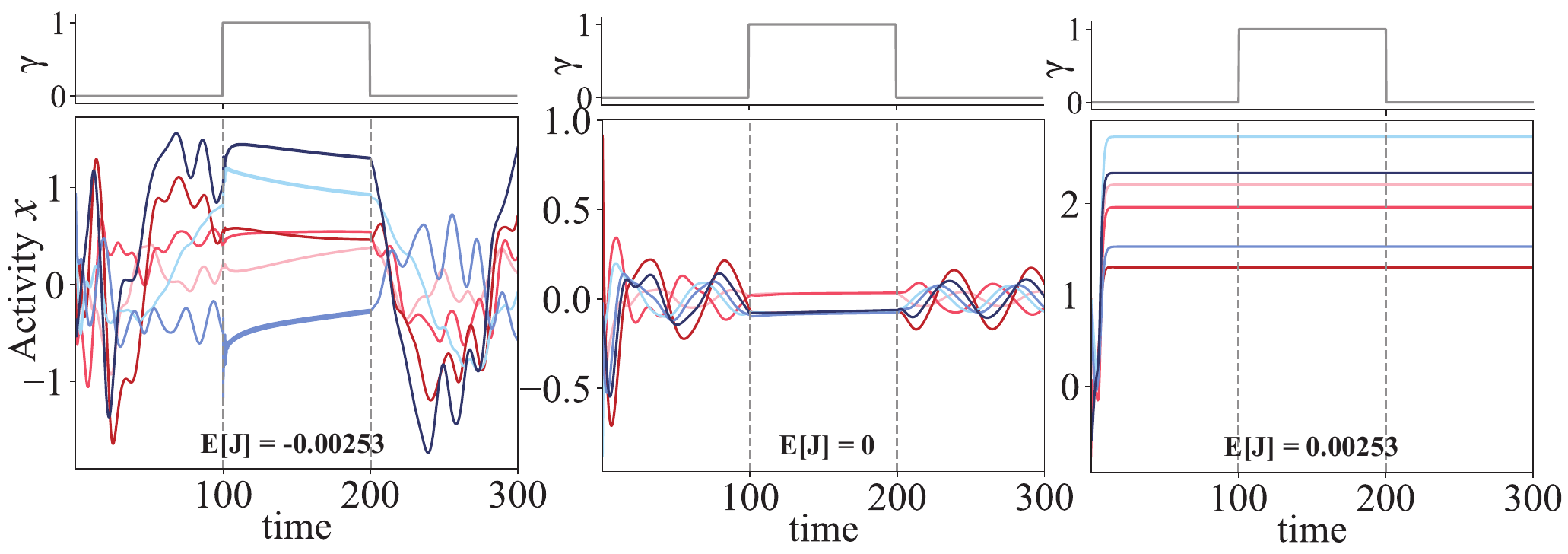}
\caption{Dynamics of EI neural networks  ($N=1\,000$). (Left) Inhibitory dominated network with $\mathbb{E}[J]<0$. (Middle) Structurally balanced network with vanishing
mean. (Right) Excitatory dominated network with positive mean. These three cases are in one-to-one correspondence with those in Fig.~\ref{fig5}. } 
\label{fig6}
\end{figure}

We can thus define the structurally balanced setting as $\mathbb{E}[J]=0$, while a positive (negative) mean implies the excitatory (inhibitory) dominated setting. According to
our above construction, it is straightforward to derive the explicit expression for the mean:
\begin{equation}
\mathbb{E}[J]=\alpha[f\tilde{\mu}_{\rm e}+(1-f)\tilde{\mu}_{\rm i}].
\end{equation}
This expectation determines the eigenvalue outlier of the sparse random matrix that obeys Dale's law. We can also calculate the radius of the eigen-spectrum disc as
$\mathcal{R}=\sqrt{N[f\sigma_{s\mathrm{e}}^2+(1-f)\sigma_{s\mathrm{i}}^2]}$, where $\sigma_{sl}^2=\alpha(1-\alpha)\tilde{\mu}^2_l+\alpha\tilde{\sigma}_l^2$ ($l=\mathrm{e},\mathrm{i}$). 
Figure~\ref{fig5} shows three examples of the connectivity matrix and the associated coupling distributions.

Given the constructed excitatory-inhibitory (EI) networks, we apply our OR-regulated dynamics to the biologically realistic systems and find that the dynamics can still be controlled.
For example, in the case of structurally balanced networks, the neural dynamics can be slowed by turning on the OR term, while shutting down the OR term can accelerate 
the dynamics with strong chaotic fluctuations. However, if the vanilla dynamics is not chaotic, e.g., a non-trivial fixed point, then we find that the OR term seems to play no role (see the right panel of Fig.~\ref{fig6}).

\subsection{Memory and predictive processing performance}
\begin{figure}
\centering
\includegraphics[width=0.65\textwidth]{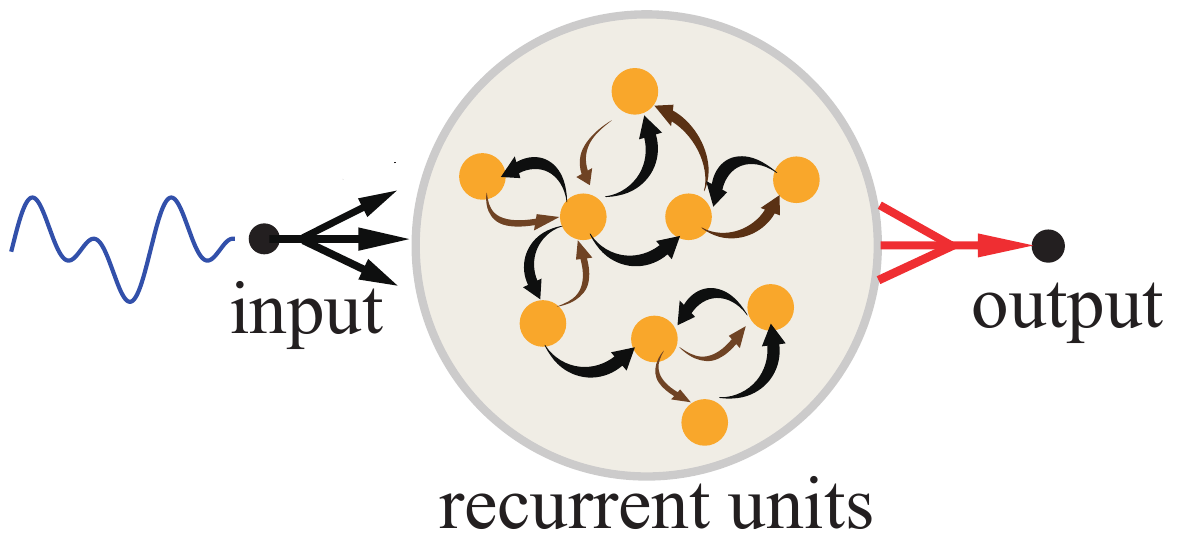}
\caption{Sketch of the reservoir computing for the stimuli forecasting and recalling experiments. The reservoir neurons follow the dynamics:
$\dot{\mathbf{x}}=-\mathbf{x}+\mathbf{J}\phi(\mathbf{x})+\mathbb{S}(t)\mathbf{u}$, where $\mathbf{u}$ is an all-one vector. Only the readout weight $\mathbf{W}$ is trained to yield
the prediction $\hat{z}(t)=\sum_iW_ix_i(t)$ compared with the target (specified in the main text).}
\label{fig7a}
\end{figure}
The vanilla RNN can be used for information processing, e.g., working memory, decision making, and motor control~\cite{Sus-2009,Zou-2023}. The original random couplings in the neural pool
are not trained, and only the linear readout weight is trained (see an illustration in Fig.~\ref{fig7a}), which is also called reservoir computing~\cite{RC-2001}. Here, we will test the gradient neural dynamics yet with
asymmetric couplings on two computational tasks. The first one is recalling the previous stimuli based on the current neural state. We thus define the temporal separation as
$\tau$. It is expected that the task difficulty grows with $\tau$. The second task is forecasting a future event, i.e., sequence anticipation. The computational ability in both
tasks is a core component underlying intelligent behavior~\cite{Eric-2021,Seq-2023,PD-2020}. To have a quantitative measure of the network performance, we define the following memory function~\cite{Memory-2012}:
\begin{equation}\label{memory}
m(\tau)=1-\frac{\min_{W}\epsilon_W(\tau)}{\langle\mathbf{z}^2\rangle},
\end{equation}
where $\langle\cdot\rangle$ denotes the temporal average, $W$ denotes the readout weight (an $N$-dimensional real-valued vector), and $\epsilon_W(\tau)$ is a $W$-dependent mean-squared
error between the actual reconstruction (or prediction) and the target $\mathbf{z}$. The explicit expression of $\epsilon_W(\tau)$ is given by $\epsilon_W(\tau)=\langle\|\hat{\mathbf{z}}-\mathbf{z}\|^2\rangle$, where the linear readout reads $\hat{z}(t)=\mathbf{W}^\top\mathbf{x}(t)$, and the average is a temporal average. 

We assume the external stimuli can be written as $\mathbb{S}(t)$. If $z(t)=\mathbb{S}(t-\tau)$, Equation \eqref{memory} becomes 
a memory function, i.e., measuring how well the network reconstructs the previous input. If $z(t)=\mathbb{S}(t+\tau)$, $m(\tau)$ measures how well the network predicts future stimuli. Given a time window $t\in[0,L]$, we concatenate the neural states into a matrix $\mathbf{X}\in\mathbb{R}^{L\times N}$, and the target becomes accordingly a vector of length $L$. Therefore,
the prediction $\hat{\mathbf{z}}=\mathbf{X}\mathbf{W}$. To minimize the mean-squared error, we have the following optimal solution by setting the derivative zero:
\begin{equation}
\mathbf{W}^*=(\langle\mathbf{X}^\top\mathbf{X}\rangle)^{-1}\langle\mathbf{X}^\top\mathbf{z}\rangle,
\end{equation}
where we assume that $L>N$ and such matrix inverse is guaranteed to exist, and otherwise $\alpha\id_{N}$ ($\alpha>0$) can be added to the covariance matrix.

\begin{figure}
\centering
\includegraphics[width=\textwidth]{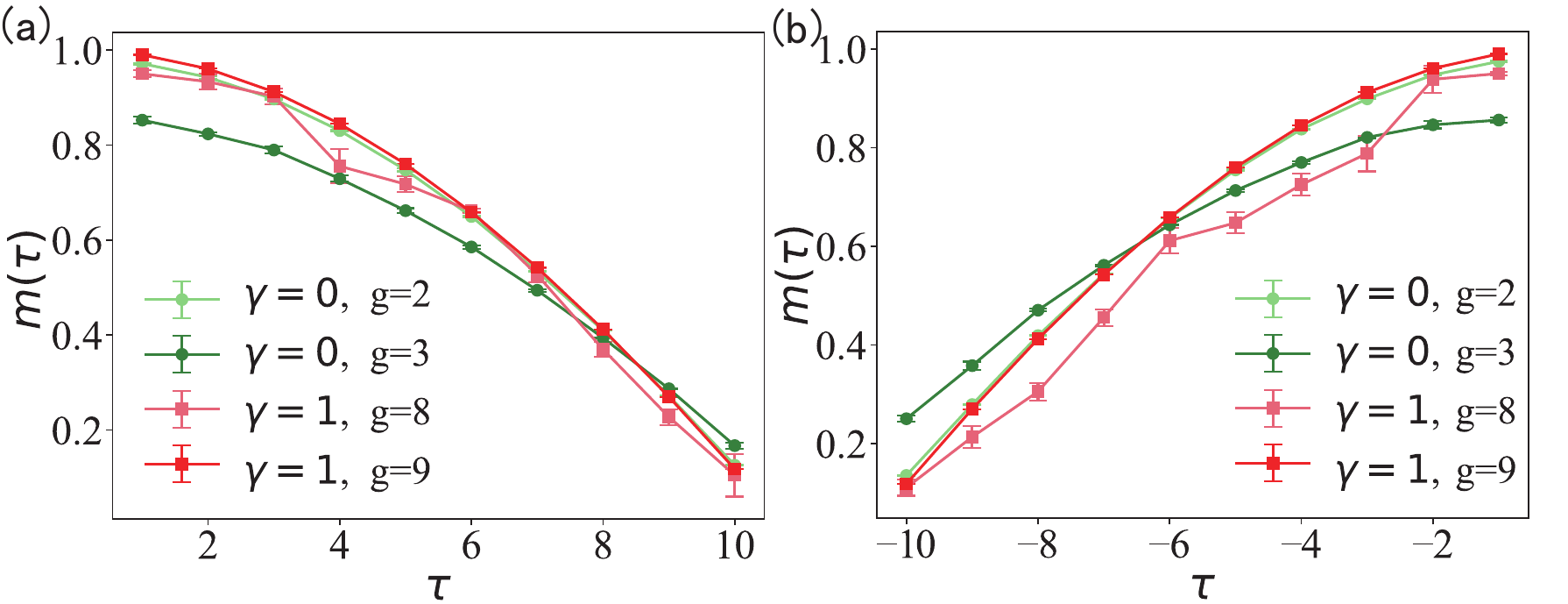}
\caption{Memory or prediction curve for the gradient dynamics with asymmetric couplings ($N=1\,000$). The stimuli $\mathbb{S}(t)=\sin(\omega_1t)+\sin(\omega_2t)$, where
$\omega_1=0.02\pi$ and $\omega_2=0.04\pi$. (a) Prediction curve. The stimulus at time $t+\tau$ is forecasted from the neural activity at the time step $t$. (b) Memory curve. The stimulus at time $t-\tau$ is recalled
from the neural activity at the time step $t$. 
The vanilla RNN dynamics without the OR term are compared in these experiments. The error bar characterizes the fluctuation across five independent experiments.} 
\label{fig7}
\end{figure}

Given the optimal solution $\mathbf{W}^*$, one can prove that $\langle\hat{\mathbf{z}}^\top\mathbf{\hat{z}}\rangle=\langle\mathbf{\hat{z}}^\top\mathbf{z}\rangle$. Therefore, the memory function can be simplified as
\begin{equation}
m(\tau)=\frac{\langle\mathbf{\hat{z}}^\top\mathbf{\hat{z}}\rangle}{\langle\mathbf{z}^2\rangle}=\frac{\langle\mathbf{z}^\top\mathbf{X}\rangle(\langle\mathbf{X}^\top\mathbf{X}\rangle)^{-1}\langle\mathbf{X}^\top\mathbf{z}\rangle}{\langle\mathbf{z}^2\rangle},
\end{equation}
where $\langle\cdot\rangle$ denotes a temporal average. In the following experiments, we assume $\mathbb{S}(t)=\sin(\omega_1t)+\sin(\omega_2t)$, where $\omega_1$ and $\omega_2$ are two different angular frequencies of the signal. Results are shown in Fig.~\ref{fig7}. It is evident that when $g\geq8$, the gradient
neural dynamics with the OR term are comparable to the vanilla non-gradient dynamics in both retrospective and prospective computations. The former becomes even more superior than the latter especially when the time separation is short. 

\section{Summary}

In this paper, we study the collective behavior of gradient neural dynamics, which is derived from vanilla non-gradient RNN by keeping the kinetic energy decreasing over dynamics.
We show that an additional OR term in the gradient neural dynamics can be used to freeze the strong chaotic fluctuation in the vanilla dynamics, and therefore the freezing without synaptic plasticity can serve as a sort of working memory for information processing. The OR term is able to make the dynamics approach the unstable fixed points in the vanilla dynamics. The number of these fixed points is revealed to grow exponentially with the number of degrees of freedom in the system~\cite{PRL-2013,Wang-2024}. Moreover, we also verified that this kind of freezing can also occur in more biologically realistic networks, such as networks composed of excitatory and inhibitory neurons. The gradient neural dynamics at a larger synaptic gain parameter can also be useful for memory and predictive processing. The OR-regulated dynamics are thus an alternative way to freeze chaos, in addition to the commonly adopted Hebbian plasticity. However, it remains unclear how a real neural circuit structure can implement this sort of freezing chaos without synaptic plasticity. Furthermore, it would be interesting in future works to use the gradient neural dynamics with asymmetric couplings to store information, as fixed points can be approached. We will come to these points in forthcoming works.

\section*{Acknowledgments}
We thank Yu Zhendong for helpful discussions. This research was supported by the National Natural Science Foundation of China for
Grant number 12475045, and Guangdong Provincial Key Laboratory of Magnetoelectric Physics and Devices (No. 2022B1212010008), and Guangdong Basic and Applied Basic Research Foundation (Grant No. 2023B1515040023).

\begin{thebibliography}{10}

\bibitem{Chaos-1988}
H.~Sompolinsky, A.~Crisanti, and H.~J. Sommers.
\newblock Chaos in random neural networks.
\newblock {\em Phys. Rev. Lett.}, 61:259--262, 1988.

\bibitem{Pham-2025}
Tuan~Minh Pham, Albert Alonso, and Karel Proesmans.
\newblock Irreversibility in non-reciprocal chaotic systems.
\newblock {\em New Journal of Physics}, 27(2):023003, 2025.

\bibitem{Qiu-2024}
Junbin Qiu and Haiping Huang.
\newblock An optimization-based equilibrium measure describing fixed points of
  non-equilibrium dynamics: application to the edge of chaos.
\newblock {\em Communications in Theoretical Physics}, 77(3):035601, 2025.

\bibitem{Huang-2024}
Haiping Huang.
\newblock Eight challenges in developing theory of intelligence.
\newblock {\em Front. Comput. Neurosci}, 18:1388166, 2024.

\bibitem{Kadmon-2024}
Jonathan Kadmon.
\newblock Efficient coding with chaotic neural networks: A journey from
  neuroscience to physics and back.
\newblock {\em 2408.01949}, 2024.

\bibitem{Clark-2024}
David~G. Clark and L.~F. Abbott.
\newblock Theory of coupled neuronal-synaptic dynamics.
\newblock {\em Phys. Rev. X}, 14:021001, 2024.

\bibitem{PRL-2013}
Gilles Wainrib and Jonathan Touboul.
\newblock Topological and dynamical complexity of random neural networks.
\newblock {\em Phys. Rev. Lett.}, 110:118101, 2013.

\bibitem{Huang-2022}
Haiping Huang.
\newblock {\em Statistical Mechanics of Neural Networks}.
\newblock Springer, Singapore, 2022.

\bibitem{Huang-2014}
Haiping Huang and Yoshiyuki Kabashima.
\newblock Dynamics of asymmetric kinetic ising systems revisited.
\newblock {\em Journal of Statistical Mechanics: Theory and Experiment},
  2014(5):P05020, 2014.

\bibitem{AP-2004}
P~Ao.
\newblock Potential in stochastic differential equations: novel construction.
\newblock {\em Journal of Physics A: Mathematical and General}, 37(3):L25,
  2004.

\bibitem{WJ-2013}
Han Yan, Lei Zhao, Liang Hu, Xidi Wang, Erkang Wang, and Jin Wang.
\newblock Nonequilibrium landscape theory of neural networks.
\newblock {\em Proceedings of the National Academy of Sciences of the United
  States of America}, 110(45):E4185--E4194, 2013.

\bibitem{Yu-2025}
Zhendong Yu and Haiping Huang.
\newblock Nonequilbrium physics of generative diffusion models.
\newblock {\em Phys. Rev. E}, 111:014111, 2025.

\bibitem{JC-2003}
Julien~Clinton Sprott.
\newblock {\em {Chaos and Time-Series Analysis}}.
\newblock Oxford University Press, 2003.

\bibitem{chaos-2009}
Massimo Cencini, Fabio Cecconi, and Angelo Vulpiani.
\newblock {\em Chaos:From Simple Models to Complex Systems}.
\newblock WORLD SCIENTIFIC, Singapore, 2009.

\bibitem{ND-2014}
Wulfram Gerstner, Werner~M. Kistler, Richard Naud, and Liam Paninski.
\newblock {\em Neuronal Dynamics: From Single Neurons to Networks and Models of
  Cognition}.
\newblock Cambridge University Press, United Kingdom, 2014.

\bibitem{PRR-2023}
Isabelle~D. Harris, Hamish Meffin, Anthony~N. Burkitt, and Andre D.~H.
  Peterson.
\newblock Effect of sparsity on network stability in random neural networks
  obeying dale's law.
\newblock {\em Phys. Rev. Res.}, 5:043132, 2023.

\bibitem{Sus-2009}
David {Sussillo} and L.F. {Abbott}.
\newblock Generating coherent patterns of activity from chaotic neural
  networks.
\newblock {\em Neuron}, 63(4):544--557, 2009.

\bibitem{Zou-2023}
Wenxuan Zou, Chan Li, and Haiping Huang.
\newblock Ensemble perspective for understanding temporal credit assignment.
\newblock {\em Phys. Rev. E}, 107:024307, Feb 2023.

\bibitem{RC-2001}
Herbert Jaeger.
\newblock The “echo state” approach to analysing and training recurrent
  neural networks-with an erratum note.
\newblock {\em Bonn, Germany: German national research center for information
  technology gmd technical report}, 148(34):13, 2001.

\bibitem{Eric-2021}
Stefano Recanatesi, Matthew Farrell, Guillaume Lajoie, Sophie Deneve, Mattia
  Rigotti, and Eric Shea-Brown.
\newblock Predictive learning as a network mechanism for extracting
  low-dimensional latent space representations.
\newblock {\em Nature Communications}, 12(1):1417, 2021.

\bibitem{Seq-2023}
Matteo Saponati and Martin Vinck.
\newblock Sequence anticipation and spike-timing-dependent plasticity emerge
  from a predictive learning rule.
\newblock {\em Nature Communications}, 14(1):4985, 2023.

\bibitem{PD-2020}
Lukas Gonon, Lyudmila Grigoryeva, and Juan-Pablo Ortega.
\newblock Memory and forecasting capacities of nonlinear recurrent networks.
\newblock {\em Physica D: Nonlinear Phenomena}, 414:132721, 2020.

\bibitem{Memory-2012}
Joni Dambre, David Verstraeten, Benjamin Schrauwen, and Serge Massar.
\newblock Information processing capacity of dynamical systems.
\newblock {\em Scientific Reports}, 2(1):514, 2012.

\bibitem{Wang-2024}
Shishe Wang and Haiping Huang.
\newblock How high dimensional neural dynamics are confined in phase space.
\newblock {\em arXiv:2410.19348}, 2024.

\bibitem{HWZ-2025}
Weizhong Huang.
\newblock https://github.com/......, 2025.

\end{thebibliography}

\onecolumngrid
\appendix
\section{Orbit separation method}\label{app-a}
The orbit separation (OS) method is a widely used approach for calculating the maximal Lyapunov exponent, which is a key indicator of chaos in dynamical systems. The core idea is to track the exponential divergence of two initially close trajectories in phase space. The key features of the OS method include maintaining a small separation after every update, tracking divergence through continuous rescaling and averaging growth rates. We summarize the detailed procedure in Alg.~\ref{alg1}. The codes are available in our GitHub page~\cite{HWZ-2025}

\begin{algorithm}
\caption{Orbit separation method for calculating the maximum Lyapunov exponent}\label{alg1}
\begin{algorithmic}[1]
\Require Initial condition $\mathbf{x}_0 \in \mathbb{R}^N$, time increment $\Delta t$, total time length $L$, a random vector $\boldsymbol{\epsilon} \in \mathbb{R}^N$, a small constant $\delta$.
\State Initialize $\mathbf{x} = \mathbf{x}_0$ and $\mathbf{y} = \mathbf{x}_0 + \delta\boldsymbol{\epsilon}/\|\boldsymbol{\epsilon}\|_2$.
\State Set $t = 0$.
\State Initialize an empty list $\mathcal{L}$ to store the Lyapunov exponents.
\While{$t \leq L$}
    \State Evolve $\mathbf{x}$ and $\mathbf{y}$ using the dynamics of the system for a time step $\Delta t$.
    \State Compute the distance $d = \|\mathbf{y} - \mathbf{x}\|_2$.
    \State  Compute the Lyapunov exponent $\lambda_t = \ln \left( \frac{d}{\delta} \right)$.
    \State Append $\lambda_t$ to the list $\mathcal{L}$.
    \State Normalize $\mathbf{y}$ by replacing it with $\mathbf{x} + \delta \frac{\mathbf{y} -\mathbf{ x}}{d}$. 
    \State Set $t = t + \Delta t$.
    \EndWhile
\Ensure  Compute the maximum Lyapunov exponent $\lambda = \frac{1}{(L-t_0+1)\Delta t} \sum_{i=t_0}^{\vert\mathcal{L}\vert} \lambda_i$, where $t_0=2\,000$, and  the first $t_0-1$ estimations are discarded.
\end{algorithmic}
\end{algorithm}



\end{document}